\documentclass[aps,prl,showpacs,showkeys,twocolumn,preprintnumbers,				
               amsmath,amssymb,floatfix,
					citeautoscript,
					superscriptaddress]{revtex4-1}

\usepackage{hyperref}
\usepackage{amssymb, amsmath}
\usepackage{graphicx}	

\newcommand{\fc}{\mathcal{E}}
\newcommand{\ff}{\mathcal{F}}
\newcommand{\fx}{\mathcal{P}}
\newcommand{\nr}{\mathcal{N}}

\begin{document}

\preprint{}

\title{Polarization bistability and resultant spin rings in semiconductor microcavities } 




\author{D.~Sarkar}
\affiliation{Department of Physics and Astronomy, University of Sheffield, Sheffield S3 7RH, United Kingdom}
\author{S.~S.~Gavrilov}
\affiliation{Institute of Solid State Physics, RAS, Chernogolovka 142432, Russia}
\author{M.~Sich}
\affiliation{Department of Physics and Astronomy, University of Sheffield, Sheffield S3 7RH, United Kingdom}
\author{J.~H.~Quilter}
\affiliation{Department of Physics and Astronomy, University of Sheffield, Sheffield S3 7RH, United Kingdom}
\author{R.~A.~Bradley}
\affiliation{Department of Physics and Astronomy, University of Sheffield, Sheffield S3 7RH, United Kingdom}
\author{N.~A.~Gippius}
\affiliation{LASMEA, Universit$\acute{e}$ Blaise Pascal, UMR 6602 CNRS, 63177 Aubi$\grave{e}$re, France and A.M. Prokhorov
General Physics Institute, RAS, Moscow 119991, Russia}
\author{K.~Guda}
\affiliation{Department of Physics and Astronomy, University of Sheffield, Sheffield S3 7RH, United Kingdom}
\author{V.~D.~Kulakovskii}
\affiliation{Institute of Solid State Physics, RAS, Chernogolovka 142432, Russia}
\author{M.~S.~Skolnick}
\affiliation{Department of Physics and Astronomy, University of Sheffield, Sheffield S3 7RH, United Kingdom}
\author{D.~N.~Krizhanovskii}
\email[]{D.Krizhanovskii@sheffield.ac.uk}
\affiliation{Department of Physics and Astronomy, University of Sheffield, Sheffield S3 7RH, United Kingdom}

\date{\today}

\begin{abstract}
The transmission of a pump laser resonant with the lower polariton branch of a semiconductor microcavity is shown to be highly dependent on the degree of circular polarization of the pump. Spin dependent anisotropy of polariton-polariton interactions allows the internal polarization to be controlled by varying the pump power. The formation of spatial patterns, spin rings with high degree of circular polarization, arising as a result of polarization bistability, is observed. A phenomenological model based on spin dependent Gross-Pitaevskii equations provides a good description of the experimental results. Inclusion of interactions with the incoherent exciton reservoir, which provides spin-independent blueshifts of the polariton modes, is found to be essential. 
\end{abstract}


\maketitle



Nonlinear interactions in optical systems result in a variety of important phenomena such as frequency conversion, parametric oscillation, bistability,  pattern formation and self-organization. In this context hybrid light-matter particles, polaritons, which form due to strong exciton-photon coupling in semiconductor microcavities (MCs), attract much attention \cite{KavokinBook}. In this case strong nonlinear interactions due to the excitonic component of polaritons lead to stimulated polariton-polariton scattering and optical parametric oscillation \cite{Savvidis00prl, krizh2000}, bistability \cite{Tredicucci96pra, baas04pra} and superfluidity \cite{Carusotto04prl, Amo09natphys}. Bose-Einstein condensation of polariton quasi-particles has also been reported \cite{Kasprzak06nat}. It is notable that compared to weakly coupled light/matter microcavity systems, polariton nonlinear interactions are several orders of magnitude stronger \cite{KavokinBook}. 

A further distinguishing feature of polariton systems arises from their spin properties. In particular, polaritons with parallel spins repel, whereas polaritons with opposite spins attract. Such interactions provide blueshifts and redshifts respectively of the energies of coherent polariton modes. This anisotropy in spin properties results in polarization bistability and multistability predicted recently \cite{Gippius07prl}. Polariton polarization bistability has also been predicted to lead to the formation of spatial spin rings of high degree of circular polarization (DCP) \cite{Shelykh08prl}.  These non-linear spin properties and spatial patterns may lead to novel optical/spin-based devices such as fast optical modulators, spin switches \cite{Shelykh08prl, AmoNPhoton} and polariton logic elements (polariton neurons) \cite{Liew08prl}, operating at high picosecond speeds and very low pump powers.  

In the present work we investigate bistability of spin-up and spin-down polariton fields as a function of the intensity and polarization of an external pump beam. As a result of spin dependent polariton-polariton interactions \cite{Gippius07prl} we are able to switch abruptly the internal polariton DCP by 40-50\% by tuning the pump power. Despite strong photonic disorder we demonstrate the formation of spatial ring patterns of high DCP, a result of the bistable threshold-like behavior of the DCP for spatially non-uniform excitation \cite{Shelykh08prl}. The pump power behavior and the similar bistability thresholds for spin-up and spin-down coherent polariton components for unequal spin polarized internal fields, indicate the influence of the incoherent exciton reservoir in leading to blueshifts towards resonance of the polarized polariton modes. The experimental results are explained by a theoretical model based on nonlinear Gross-Pitaevskii equations, taking into account nonlinear polariton interactions and coupling to the reservoir.

\begin{figure}[t!]  
  \includegraphics[width=\linewidth]{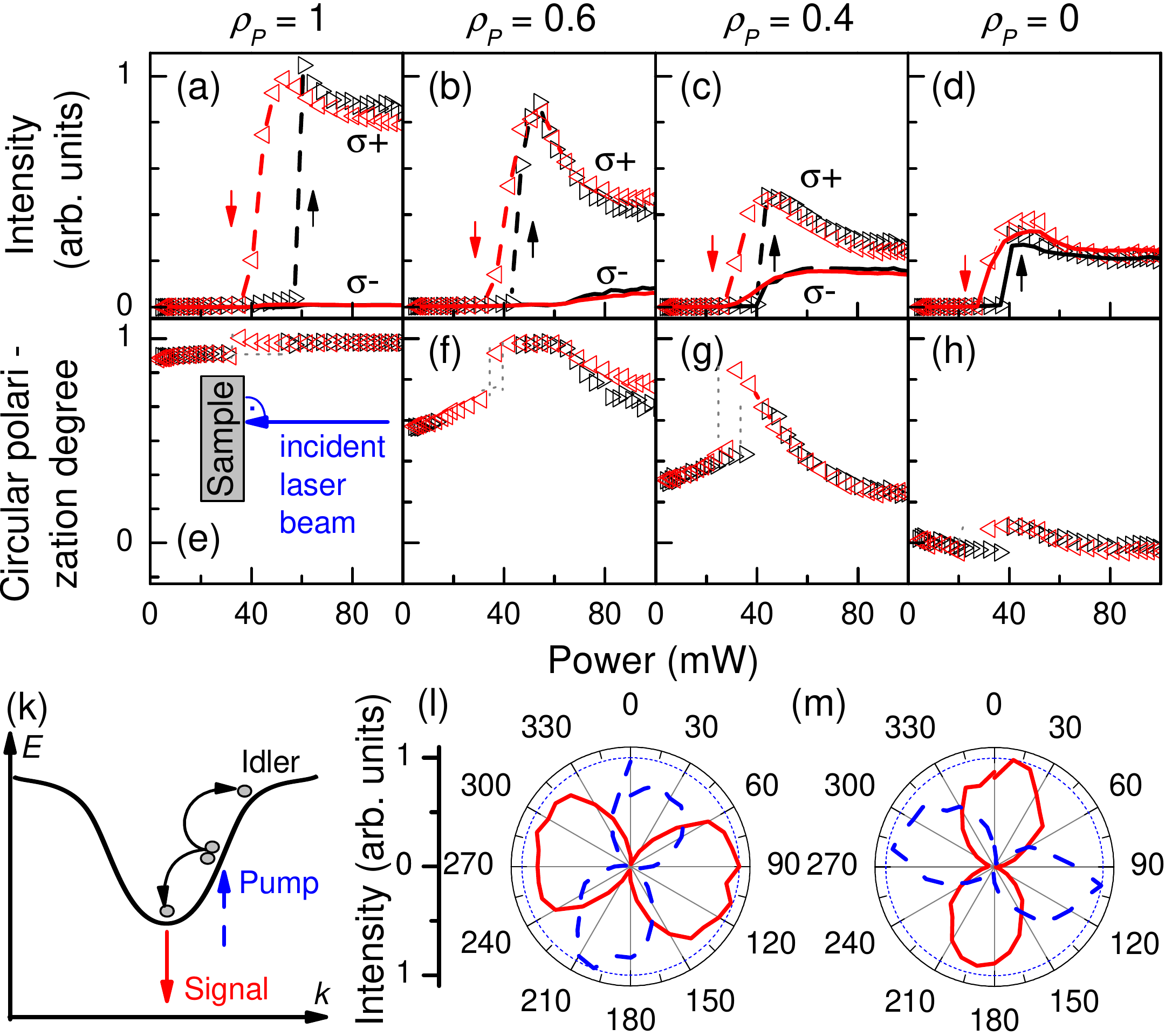}

  \caption{(Color online) (a)-(d) Normal incidence transmitted intensity versus external pump power for $\rho_P = 1$, $0.6$, $0.4$ and $0$. Triangles (lines) correspond to the intensity of the $\sigma^+$ ($\sigma^-$) component. Arrows show the direction in which the pump intensity is varied. (e)-(h): Corresponding DCP of transmission versus pump power. Grey dotted lines serve as a guide to the eye. The inset in (e) shows a schematic of the experimental geometry for (a)-(h).
	(k) Schematic diagram of OPO. 
	(l) and (m) Polar plots of normalized signal (dashed line) and the laser intensity (continuous line) 
	for TE (l) and TM (m) polarized pump. 
} 
  \label{fig:1}
\end{figure}

We study the transmission of a single-mode CW Ti:sapphire laser at normal incidence ($k=0$) [see Fig.~\ref{fig:1}(e) inset] in a GaAs-based MC described in Ref.~\onlinecite{Sanvitto2006}. The detuning between exciton and photon modes is close to zero. The energy of the laser is tuned $\sim0.7$~meV above the bottom of the lower polariton (LP) branch (854.8 nm) and $\sim2$ meV below the exciton level \cite{Tredicucci96pra, baas04pra}. The polarization of the transmitted beam is analyzed by a quarter-wave plate and a linear polarizer. Spatially resolved images were recorded by a CCD camera. The sample is mounted in a He-bath cryostat at 5 K. 

A well-known property of the interacting polariton system is bistability of transmitted intensity (or internal polariton field), which has a characteristic S-shape dependence versus pump power \cite{Tredicucci96pra, baas04pra}. It is observed at positive detuning between the pump energy and the LP mode.  At low pump power the transmission is very low, since the laser energy is out of resonance with the LP mode. With increasing pump intensity polariton-polariton interactions shift the LP energy towards resonance with the pump energy \cite{Gippius07prl} leading to an abrupt increase of the LP population and hence the transmitted intensity \cite{Tredicucci96pra, baas04pra}. When the pump power is lowered the LP energy jumps back at a lower threshold power, giving rise to a hysteresis loop of transmitted pump power dependence, i.e. bistable behavior.

Figures~\ref{fig:1}(a)-(d) show the intensities (\textit{$I(\sigma^{\pm})$}) of $\sigma^+$ (triangles) and $\sigma^-$ (lines) polarized components of the transmitted beam versus pump power ($I_P$) for various pump DCPs $\rho_P=\frac{I_P(\sigma^+)-I_P(\sigma^-)}{I_P(\sigma^+)+I_P(\sigma^-)}$. The signal is detected from a small region of $2\times2$ $\mu$m$^2$, over which the pump intensity is nearly constant.
For $\rho_P=1$, $I(\sigma^+)$ shows an abrupt switch followed by a slow decrease with increasing pump power. A hysteresis loop in the pump power dependence of $I(\sigma^+)$ is observed in accordance with polariton bistable behavior \cite{baas04pra}.  
Notably, on reducing the pump DCP from $\rho_P=1$ to $\sim0$ the bistability threshold power and the transmitted intensity at the threshold for the $\sigma^+$-component decrease by a factor of 1.5 and 3, respectively.
 
The behavior of the minority $\sigma^-$ component versus pump power is different. $I(\sigma^-)$ is zero for $\sigma^+$ circularly polarized  excitation (i.e $\rho_P=1$), but appears for smaller DCP of the laser until it approaches the same intensity as the $\sigma^+$ component for a nearly linearly polarized pump [Fig.~\ref{fig:1}(d)]. 
In the latter case, as expected both $I(\sigma^+)$ and $I(\sigma^-)$ show nearly the same threshold, intensity and hysteresis loop width. For elliptically polarized pumping at $\rho_P = 0.6$ and $0.4$, the $\sigma^-$ component emerges respectively at powers 1.5 times larger or nearly equal to the threshold of the $\sigma^+$ component. Surprisingly, $I(\sigma^-)$ exhibits a smooth intensity increase at threshold instead of a steep intensity jump and shows only a weak ($\rho_P=0.4$) or no ($\rho_P=0.6$) hysteresis behavior.
 
The corresponding pump power dependence of the DCP of the transmitted light $\rho_c=\frac{I(\sigma^+)-I(\sigma^-)}{I(\sigma^+)+I(\sigma^-)}$ is shown in Fig.~\ref{fig:1}(e)-(h). At very low pump powers in the linear regime the DCP of the pump defines $\rho_c$, which then increases slightly by $10\%$ with increasing pump power. When the threshold power is reached, $\rho_c$ jumps to nearly one for $\rho_P = 0.6$ and $0.4$. For  $\rho_P = 0$ only a slight change to $0.1$ at threshold is observed. With further increase of pump power, $\rho_c$ decreases again and eventually approaches the pump DCP $\rho_P$ at very high powers. The hysteresis behavior of $\rho_c$ follows that of the $\sigma^+$ component of the transmitted intensity, demonstrating polarization bistability. 

The fact that the $I(\sigma^+)$ dependence has 1.5 times lower threshold than $I(\sigma^-)$ at $\rho_P = 0.6$ indicates a difference between the strengths of $\sigma^+\sigma^+$ and $\sigma^+\sigma^-$ polariton-polariton 
interactions \cite{Gippius07prl}.  The sign and the strength of these interactions is strongly influenced by the 
biexciton resonance
\cite{Wouters:2007}. In particular, for pump energies $E_p$ below the biexciton level $2E_p<2E_X-E_b$ ($E_X$ and $E_b\sim 1$ meV are the exciton energy and the biexciton binding energy, respectively) an attraction (repulsion) between cross (co)-circularly polarized polaritons is expected \cite{Wouters:2007}.

Attractive coupling between coherent $\sigma^+$ and $\sigma^-$ modes is verified by studying the polarization of the stimulated emission formed at $k=0$ under  resonant excitation into the LP branch at k-vectors $k_p\sim 1.4$~$\mu$m$^{-1}$. In this case pump-pump scattering results in formation of a macroscopically occupied signal at k=0 and idler at $k=2k_p$ in an optical parametric oscillator (OPO) configuration [Fig.~\ref{fig:1}(k)].
Polar intensity diagrams of the signal (dashed line) and the pump (solid line) are shown in Fig.~1(l) and (m) for TE and TM polarized pump, respectively. In both cases the orientation of the linearly polarized stimulated emission at $k=0$ is rotated by $\sim90^\circ$ with respect to the excitation \cite{Krizhanovskii06prb}. The $\sim90^\circ$ rotation has been observed for pump energies in the broad range $1.5-3$ meV below the exciton level.  As discussed in Ref.~\onlinecite{Krizhanovskii06prb} this can only be accounted for by the attractive character of $\sigma^+\sigma^-$ interactions.

For weak $\sigma^+\sigma^-$ attractive interactions the bistability threshold of $I(\sigma^+)$ should be at least 4 and 2.3 times less than that for $I(\sigma^-)$ at $\rho_P=0.6$ and $0.4$, respectively \cite{Gippius07prl}, according to the ratio of the cross-circularly polarized pump components $I_P(\sigma^+)/I_P(\sigma^-)$ of 4 and 2.3, respectively. Similarly, the bistability threshold is expected to be at least a factor of two larger for linearly ($\rho_P=0$) than for circularly ($\rho_P=1$) polarized pumping. These expectations are in contrast to the results observed in Fig.~\ref{fig:1}, where $I(\sigma^+)$ thresholds at $\rho_P=0.6$ and $\rho_P=0.4$ are observed at factors of 1.5 and 1 below $I(\sigma^-)$, respectively, and the ratio of the threshold for $\rho_P=0$ to $\rho_P=1$ is $\sim0.7$. This instead indicates the strong role of a repulsive blue shift interaction, which shifts the $\sigma^-$ component towards resonance resulting in a rapid increase of the intensity of $\sigma^-$ when the $\sigma^+$ mode is switched on. 

The experimental results can be explained if one takes into account coupling between the coherently driven polariton modes and the incoherent excitonic reservoir, which provides the necessary blue shift. At energies below the exciton emission this reservoir consists of dark exciton states and weakly coupled localised excitons (LE), which introduce additional damping of polaritons \cite{KrizhSSC}. For the inhomogeneously broadened exciton level (FWHM$\sim1.5$ meV) in our sample the density of LE states at 2-3 meV below $E_X$ is estimated to be 3-4 orders of magnitude larger than that of polaritons \cite{KrizhSSC}.

We adapted the model in Ref.~\onlinecite{Gippius07prl} taking into account transitions of
the optically driven excitons into the incoherent reservoir in
which the overall pseudospin is relaxed. As a result, the reservoir
provides equal blueshifts for both polarized coherent spin states in addition to the attractive coupling between them which leads to redshift. This assumption is justified given that the reservoir population is unpolarized due to fast exciton spin relaxation ($\sim 20$~ps) \cite{LSham94} shorter than the reservoir lifetime ($\sim 50$~ps) \cite{KrizhSSC}.

Our model considers the intra-cavity $\sigma^{\pm}$ polarized electric field $\fc_{\pm}$  and
coherent exciton polarization $\fx_{\pm}$, respectively, coupled with the 
exciton population ($\nr$) in the incoherent reservoir:
\begin{align}
  i \dot{\fc}_{\pm} &= (\omega_\mathrm{c} - i\gamma_\mathrm{c}) \,
  \fc_{\pm} + \alpha \ff_{\pm} + \beta \fx_{\pm}, \label{ef_p} \\
  i \dot{\fx}_{\pm} &= \bigl[ \omega_\mathrm{x} + V_1 |\fx_{\pm}|^2 + V_2
  |\fx_{\mp}|^2 + (V_1 + V_2) \, \nr/2 - {} \nonumber \\
  &i \left( \gamma_\mathrm{x} + V_\mathrm{r}
    |\fx_{\mp}|^2 \right) \bigr]
  \fx_{\pm} + A \fc_{\pm}, \label{xp_p} \\
  \dot{\nr}~{} &= -\gamma_\mathrm{r}  \nr + 
  4 V_\mathrm{r} |\fx_+|^2 |\fx_-|^2. \label{res}
\end{align}
Here, $\ff$ is the incident electric field (the pump) taken as
a plane wave: $\ff_\pm \propto e^{-i\omega_\mathrm{p}t}$;
$\omega_\mathrm{c,x}$ and $\gamma_\mathrm{c,x}$ are eigenfrequencies
and decay rates of the intra-cavity photon and exciton modes; $\alpha$ describes the response of the intracavity electric field to the external pump; $A$ and $\beta$ describe the exciton-photon coupling ($2 \sqrt{A\beta}$ is equal to the Rabi splitting);
$V_{1,2}$ are matrix elements of the interaction between excitons
with the same ($V_1$) and opposite ($V_2$) circular polarizations; $V_1 > 0$, $V_2 < 0$ and $|V_2 / V_1| \ll 1$ corresponding to strongly anisotropic spin interactions \cite{Krizhanovskii06prb}. 

The transitions from the coherently driven polariton modes into the reservoir are introduced phenomenologically as
exciton non-radiative decay rates in combination with a rate equation for the reservoir
occupation (Eq.~\ref{res}). 
The term $ V_\mathrm{r} |\fx_{\mp}|^2$ is the decay rate of the $\fx_{\pm}$ component due to scattering between coherent excitons with opposite $\sigma^{\pm}$ polarizations. It provides occupation
of the reservoir at a rate $4 V_\mathrm{r} |\fx_+|^2 |\fx_-|^2$ per unit time.  
$V_\mathrm{r}$ may microscopically arise from the
scattering of two bright excitons with opposite spin ($J_z = -1$
and $J_z = +1$) to dark excitons ($J_z = -2$ and $J_z = +2$) (see Ref.~\onlinecite{Inoue:2000} for
details). It may also arise from transitions into the exciton reservoir via virtual creation of biexcitons. 
Exciton decay into the reservoir due to $\sigma^+\sigma^-$ scattering is likely to dominate over $\sigma^+\sigma^+$ scattering, since in pump-probe experiments in MCs the absorption of a $\sigma^-$ polarized probe is observed to be strongly enhanced by a factor of 3-4 for $\sigma^+$ polarized pumping \cite{KrizhSSC}.
Finally, $\gamma_\mathrm{r}$ is the exciton decay rate in the reservoir. The coefficients $\gamma_\mathrm{r}$
and $V_\mathrm{r}$ correspond to effective mean values. We do not consider
their derivation from a microscopic Hamiltonian. 

\begin{figure}[t]  
  \includegraphics[width=\linewidth]{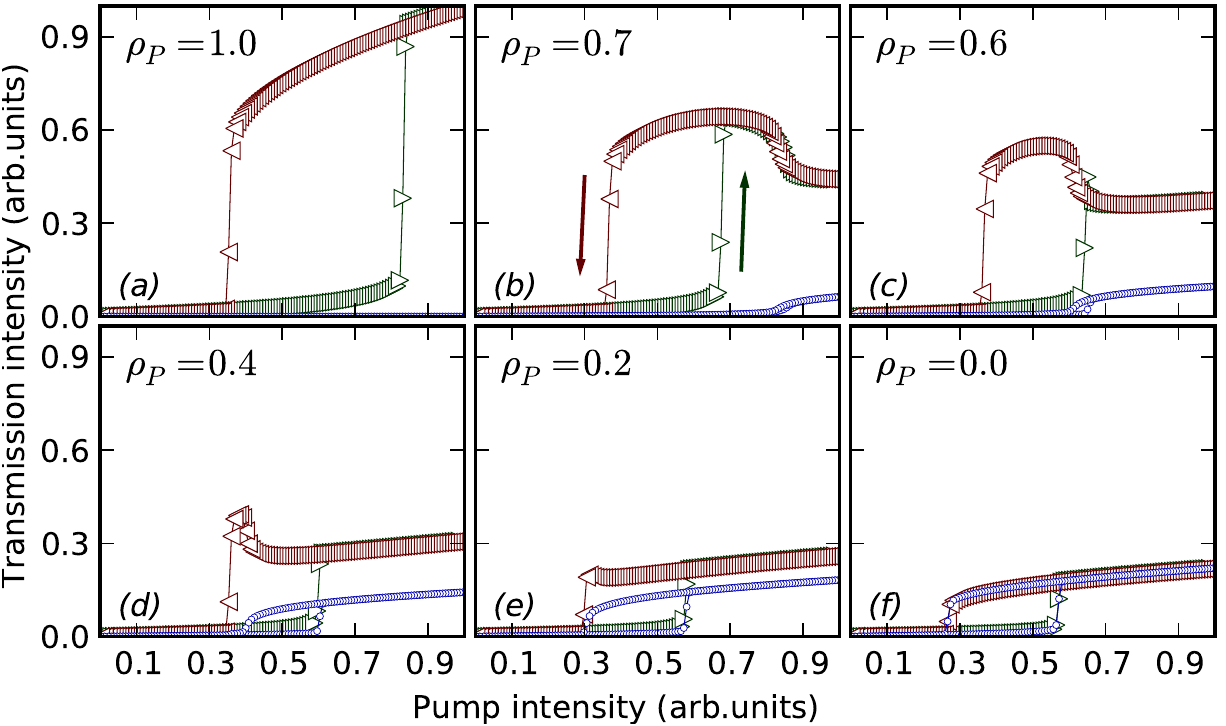}
  \caption{The calculated dependences of the cavity transmission
    intensity in two circular polarizations ($\sigma^+$ and
    $\sigma^-$) vs pump intensity, for several DCP  of
    the pump beam ($\rho$). The dominant polarization component
    ($\sigma^+$) is shown by triangles `$\vartriangleright$' and
    `$\vartriangleleft$' which correspond to the increase and decrease
    of the pump power, respectively, whereas the $\sigma^-$ intensity
    is shown by blue lines. The Rabi splitting, detuning between exciton and photon mode and detuning between laser energy and LP branch are the same as in the experiment.}
  \label{fig:theor:loops}
\end{figure}

Calculated transmission intensities versus pump power are presented in Fig.~\ref{fig:theor:loops} for various pump DCP $\rho_P$. The simulation parameters are $V_2=-0.1\cdot V_1$, 
$V_\mathrm{r} = 0.6 \cdot V_1$, $\hbar\gamma_\mathrm{r}=0.1$\,meV and $\hbar\gamma_\mathrm{x}=0.2$\,meV which is close to the measured polariton linewidth. The key results of the experiment are well reproduced. Firstly, the decrease in the $\sigma^+$ transmission with increasing pump power [Fig.~\ref{fig:theor:loops}(b)-(d)] above threshold is observed, which arises from the onset of the nonlinear decay ($V_\mathrm{r}$ term) of polaritons as the minor
$\sigma^-$ polariton component is populated.  In turn, this nonlinear decay leads to damping of the $\sigma^-$ component
and hence to its smooth increase with pump power without hysteresis at $\rho>0.6$ [Fig.~\ref{fig:theor:loops}(b)-(c)]. Secondly, the bistability thresholds of both $\sigma^+$ and $\sigma^-$ components are close to each other within 20\% at all $\rho_P$ and the maximum intensity of ($\sigma^-$) component decreases with increasing $\rho_P$. Importantly, despite the attractive $\sigma^+\sigma^-$ coupling ($V_2<0$) the bistability threshold is 1.5 times higher for circularly ($\rho=1$) than linearly ($\rho=0$) polarized pumping as observed experimentally. This behavior is due to more efficient population of the reservoir by $\sigma^+\sigma^-$ interactions under linearly than circularly polarized pumping, which blueshifts the LP modes and hence reduces the bistability threshold at $\rho=0$. 

We note that if we do not take into account the effect of the reservoir and assume instead repulsive interaction between cross-polarized polaritons in contradiction to the results of Fig.~\ref{fig:1}(f) and (g) we cannot reproduce simultaneously the observed strong jumps of DCP by 40$\%$ with pump power (Fig.~\ref{fig:1}) and the ratio of bistability thresholds for the cases of circularly and linearly polarized pumps.

\begin{figure}[t!]  
\includegraphics[width=\linewidth]{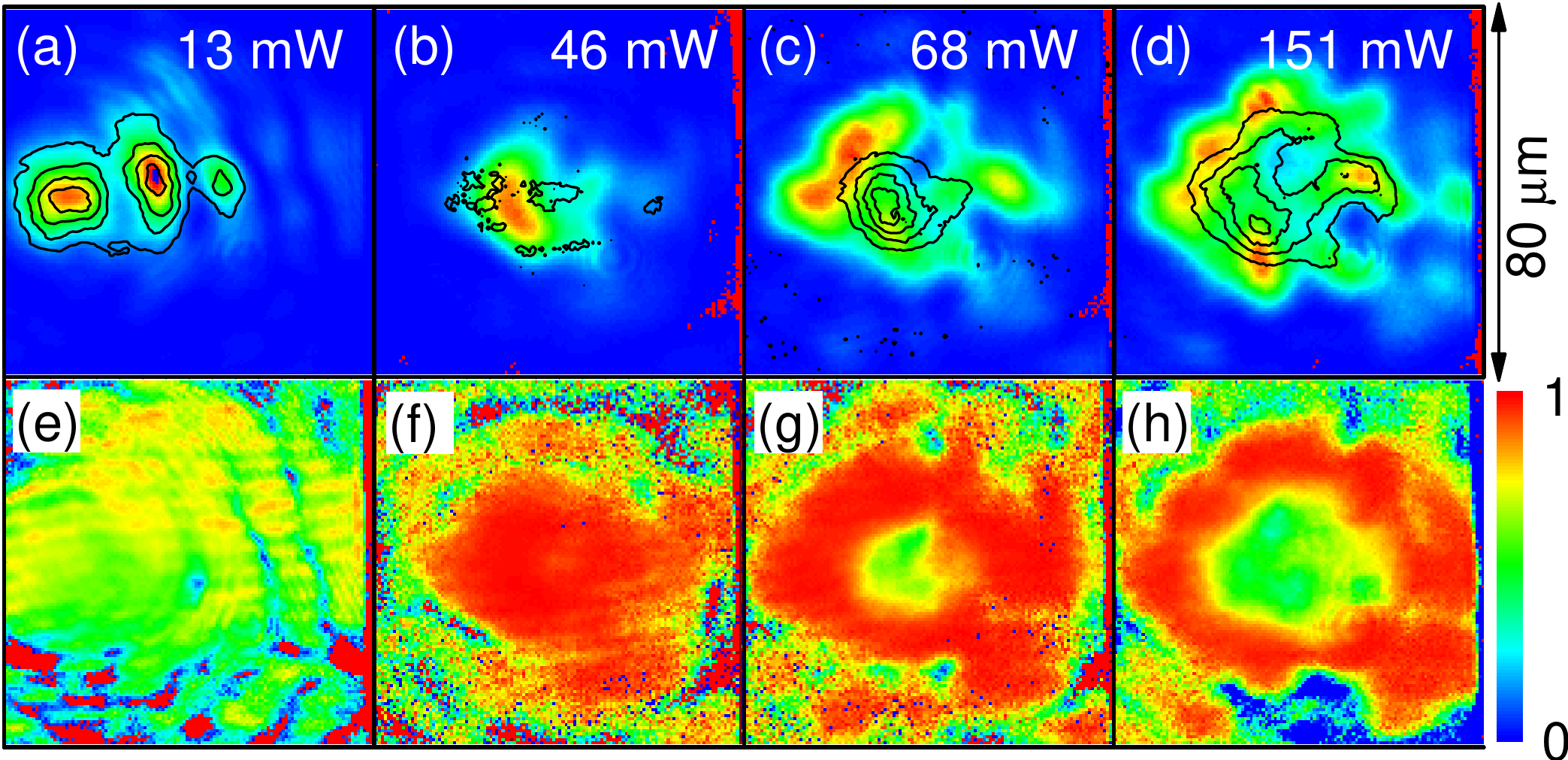}				
  \caption{(Color online)  Real space images of intensity (a)-(d) and corresponding circular polarization degree (e)-(h) of the transmitted beam for elliptically ($\rho_P = 0.6$) polarized pump beam at different pump powers. (a)-(d) The images in (a)-(d) show the normalized intensity of the $\sigma^+$ component, while the $\sigma^-$ component is indicated by contour lines.
 } 
  \label{fig:spin-ring}
\end{figure}

In the final part we show that the polarization bistability [Fig.~\ref{fig:1}(f) and (g)] leads to the formation of characteristic spatial ring patterns of high DCP (100 \%), which have been predicted in Ref.~\onlinecite{Shelykh08prl}. Spin rings arise due to variation of the pump power across the excitation region, resulting in polarization bistability thresholds being reached at higher total pump power near the edge of the spot than in the center. 

Figures~\ref{fig:spin-ring}(a)-(d) show spatial images of the transmitted beam at $\rho_P = 0.6$ for a series of pump powers. Note, that below threshold [Fig.~\ref{fig:spin-ring}(a)] both $\sigma^+$ and $\sigma^-$ polarized images are identical and strongly modulated by disorder potential \cite{Sanvitto2006}. With increasing power the LP mode in the central area of the spot is blueshifted into resonance with the pump allowing a large intensity of the beam to be transmitted. As the power is increased more the region where the threshold is achieved expands. Note, that although the spatial images above threshold [Figs.~\ref{fig:spin-ring}(c)-(d)] are still distorted by the disorder they have maxima away from the center of the spot at high power as expected for the case of bistability in a homogeneous sample \cite{Sanvitto2006}. This arises since the polariton blueshift above threshold is about $\sim0.7$ meV, which is larger than the amplitude of the disorder potential ($\pm0.1$ meV), hence partially screening polariton energy fluctuations. 

Figures~\ref{fig:spin-ring}(e)-(g) show the corresponding 2D spatial structure of the DCP. Below threshold (13~mW) the DCP is about 0.6, given by the pump DCP. Just above threshold (46~mW) the DCP jumps to unity in the middle of the pump spot forming a disk of high DCP [Fig.~\ref{fig:spin-ring}(f)]. With increasing pump power the DCP in the centre of the spot decreases down to 0.5, when the $\sigma^-$ mode switches on [Fig.~\ref{fig:1}(b)]. At the same time DCP jumps from 0.6 to one in a region forming a ring around the spot center [Fig.~\ref{fig:spin-ring}(g)], where the bistability threshold is achieved for the $\sigma^+$ but not the $\sigma^-$ mode. This ring of high DCP (spin ring) expands with pump power, as the $\sigma^+$ component switches on at larger distances away from the center of the spot [Fig.~\ref{fig:spin-ring}(h)]. Because the potential disorder is partially screened the formation of the DCP rings [Fig.~\ref{fig:spin-ring}(f)-(g)] is observed as expected for a pure Gaussian spot \cite{Shelykh08prl}. This is consistent with the simulations of Ref.~\onlinecite{Shelykh08prl}, predicting well-defined spin rings even in a disordered system.  

In summary, we have observed polarization bistability and spatial spin patterns as a result of spin anisotropy of polariton-polariton interactions. The experimental results are described by a phenomenological model taking into account coherent macroscopically occupied modes and an incoherent reservoir. Finally, we note that while we were preparing our manuscript experimental results similar to those in Fig.~\ref{fig:1} were published online in Ref.~\onlinecite{Paraiso}. Spatial spin patterns and the effect of the reservoir, two of the main points of our paper, were though unexplored since the studies were conducted on polaritonic dots in a sample with exciton linewidth $\sim0.5$ meV~\cite{Paraiso}.

This work was supported by EPSRC Grants EP/G001642, EP/E051448 and by RFBR of the RAS.

\begin{thebibliography}{}

\bibitem{KavokinBook} A. Kavokin et al.,
Microcavities (Oxford University Press, 2007).

\bibitem {Savvidis00prl} P. G. Savvidis et al., Phys. Rev. Lett. 84, 1547 (2000)


\bibitem{krizh2000} D. N. Krizhanovskii \textit{et al.}, Phys. Rev. B {\bf 77}, 115336 (2008) and references there in.

\bibitem {Tredicucci96pra} A. Tredicucci et al., Phys. Rev. A 54, 3493 (1996) 

\bibitem {baas04pra} A. Baas et al, Phys. Rev. A 69, 023809 (2004) 

\bibitem {Carusotto04prl} I. Carusotto and C.Ciuti Phys. Rev. Lett. 93, 166401 (2004)





\bibitem {Amo09natphys} A. Amo et al, Nature
457, 291 (2009)

\bibitem{Kasprzak06nat} J. Kasprzak et al.,
Nature {\bf 443}, 409 (2006),

\bibitem {Gippius07prl} N. A. Gippius et al, Phys. Rev. Lett. 98,
236401 (2007)

\bibitem {Liew08prl} T. C. H. Liew et al, Phys.
Rev. Lett. 101, 016402 (2008)

\bibitem {Shelykh08prl} I. A. Shelykh et al, Phys.
Rev. Lett. 100, 116401 (2008)

\bibitem {AmoNPhoton} A. Amo et al, Nature Photonics 4, 361-366 (2010) 

\bibitem {Sanvitto2006} D. Sanvitto et al., Phys. Rev. B 73, 241308 (2006) 


\bibitem {Krizhanovskii06prb} D. N. Krizhanovskii et al,  Phys.
Rev. B 73, 073303 (2006); P. Renucci et al., Phys. Rev. B 72, 075317 (2005);

\bibitem {Wouters:2007} M. Wouters, Phys. Rev. B 76, 045319 (2007).

\bibitem {KrizhSSC} D. N. Krizhanovskii et al,  Sol. State. Comm. 119 (7): 435-439 (2001) 

\bibitem {LSham94} A. Vinatierri et al, Phys.Rev.B, 50, 10868 (1994)

\bibitem {Inoue:2000} J.I. Inoue, T. Brandes, and A. Shimizu, Phys. Rev. B
61, 2863 (2000).

\bibitem {Paraiso} T. K. Paraiso et al., published online 4 July 2010 DOI: 10.1038, Nature Materials 2787

\end{thebibliography}
{}

\end{document}